\documentclass[epjCONF, onecolumn]{svjour}
\usepackage{graphics}
\usepackage{natbib}
\usepackage{epsfig}
\usepackage[varg]{txfonts} 
\usepackage[latin1]{inputenc}

\def\afe{[\alpha/\hbox{Fe}]}
\def\dex{\,{\rm dex}}
\def\kpc{\,{\rm kpc}}
\def\feh{\hbox{[Fe/H]}}
\def\kms{\,{\rm km}\,{\rm s}^{-1}}
\def\mgfe{[\hbox{Mg}/\hbox{Fe}]}
\def\figref#1{Fig.~\ref{#1}}

\session-title{Assembling the Puzzle of the Milky Way}
\begin{document}
\title{What Velocities and Eccentricities tell us about Radial Migration}
\author{Ralph Sch\"onrich\thanks{\email{rasch@mpa-garching.mpg.de}}}
\institute{Max Planck Institute for Astrophysics, Karl-Schwarzschild-Str. 1, 85741 Garching, Germany}
\abstract{
This note attempts to interpret some of the recent findings about a downtrend in the mean azimuthal velocity of low $\afe$ thin disc stars with increasing metallicity. The presence of such a trend was predicted in the model of \cite{SBI}, albeit with a slightly steeper slope. We show that in a simple picture a Galactic disc without mixing in angular momenta would display an exceedingly steep trend, while in the case of complete mixing of all stars the trend has to vanish. The difference between model and observational data can hence be interpreted as the consequence of the radial abundance gradient in the model being too high resulting in an underestimate of the migration strength. We shortly discuss the value of eccentricity distributions in constraining structure and history of the Galactic disc.
}
\maketitle
\section{About the trend in azimuthal velocities}\label{sec:Vtrend}

Among the various links between chemistry and kinematics in the solar neighbourhood is the trend in mean azimuthal velocity of stars with increasing metallicity for objects belonging chemically to the thin disc. The observational evidence for this trend can be traced back to \cite{Haywood08}, who found that metal-poor stars at low $\afe$ in local samples have large angular momentum.

As laid out in \cite{SBII} the low $\afe$ stars constitute the younger part of the Galactic disc \citep[cf.][]{Matteucci90}. The \cite{SBI} model cannot make a firm prediction about such trends for the high $\afe$ stars that formed at early epochs. It is still too uncertain what the radial abundance gradient and the radial behaviour of star formation approximately looked like. In particular either an inverse radial metallicity gradient at early times or different setups of inside-out formation can give rise to an inverse (i.e. positive) gradient of mean azimuthal velocity with $\feh$ for the older populations. We reserve the discussion of such possibilities to a paper in progress. Yet for the younger, low $\afe$ stars very firm predictions are available \citep[for a discussion of these effects in an N-Body model see][]{Loebman11}. Observations imply a very flat age-metallicity relation during the past couple of Gyrs. Hence the links between chemistry and kinematics are dominated by the radial abundance gradient in the interstellar medium: With increasing age stellar populations acquire random energy, which brings them on more eccentric orbits letting them oscillate further away from the circular orbits defined by their angular momentum. We call this "blurring". As angular momentum is conserved in an axisymmetric potential, we see in the solar neighbourhood stars from inner radii at low azimuthal velocities and vice versa, so that the radial abundance gradient gives rise to a significant downtrend of mean azimuthal velocity with metallicity. 

Radial migration shuffles stars in angular momentum, a process called "churning", but the populations should still contain some residual information on where they were born: as stars from the inner Galaxy get scattered to larger angular momenta from where they can visit the Solar neighbourhood, the metal-rich population is still preferentially concentrated towards the inner Galactocentric radii, while the metal-poor outer disc populations are preferentially at large angular momentum. The explanation in \cite{SBII} focused on the fact that radial migration is largely responsible for us being able to find those migrated objects in large numbers at all and on that we can observe this gradient in contrast to classical expectation where the increasing asymmetric drift with age should have given an inverse gradient from an sloping age-metallicity relation. Unfortunately it evoked the widespread misconception that radial migration is responsible for the downtrend of azimuthal velocities with metallicity. This is not true, because in fact radial migration tends to weaken the observed trend.

We can illustrate this by two extreme cases: Assume a radial abundance gradient of $-0.06 \dex/\kpc$ \citep[see][]{Luck11} in the young disc. In the first case we assume that stars do not change their initial angular momentum and that there is insignificant scatter of stellar abundances at birth at a certain Galactocentric radius $R_g$. We can hence write for the young population:
\begin{equation}
\feh(R_g) = \feh_0 + \frac{d\feh}{dR_g} (R_g - R_0)
\end{equation}
where $R_0 = 8.2$ and $\feh_0$ denote the Solar galactocentric radius and the local metallicity.
Assuming a disc with constant circular velocity of $V_c = 233 \kms$ we can then replace the galactocentric radius $R_g$ using angular momentum conservation and resolve the linear relationship for the local azimuthal velocity $V$:
\begin{equation}
V_\phi(\feh) = V_c\frac{R_g}{R_0} = V_c \left(1 + \frac{R_g-R_0}{R_0} \right) = V_c \left(1 + \frac{\feh - \feh_0}{R_0}\frac{dR_g}{d\feh}\right)
\end{equation}
and for the chosen parameters we obtain the slope in $V_\phi$ (on the quite meagre baseline in metallicity that would be observable):
\begin{equation}
\frac{dV_\phi}{d\feh} = \frac{V_c dR_g}{R_0 d\feh} = -474 \kms / \dex  .
\end{equation}
The result would be about $dV_\phi / d\feh \sim -300 \kms / \dex$ for the values used in \cite{SBI}, largely due to the steeper metallicity gradient assumed there. Measurement errors in metallicity would blur out the baseline and reduce the measured gradient, but this will not remove the stark contrast to what is observed in the data, as well as contributions from age-dependent evolution of the Galactic disc can hardly remove all difference on the super-solar metallicities.
Radial migration with its redistribution in angular momentum, however, does a huge change to this relation: If we assume the other extreme case of complete radial mixing, the gradient will vanish, as the distribution of populations does not depend any more on where the stars have been born. So we see that in general radial migration reduces the gradient and is both detectable by the expansion of the baseline in metallicity and by the reduced gradient. The presence of a weak velocity-metallicity correlation in data confirms that there is considerable radial migration, but the mixing is not complete.

\begin{figure}
\begin{center}
\epsfig{file=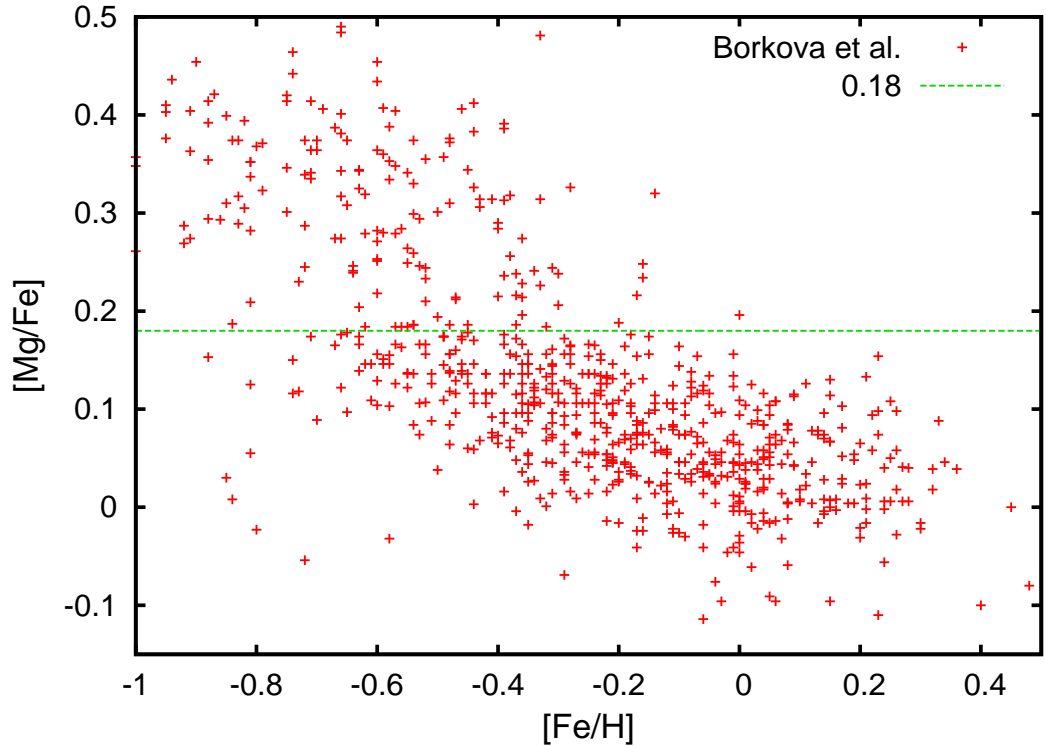, angle=-90, width=\hsize}
\caption[The metallicity plane of the Borkova et al. sample]{The metallicity plane in $\mgfe$ versus $\feh$ of the Borkova et al. sample. The depopulated region in the data advises a cut somewhere between $\mgfe = 0.14$ and $\mgfe = 0.18$ where we set the cut.}
\label{fig:Borkoverview}
\end{center}
\end{figure}  

Let us examine those trends on a real data set. From the model of \cite{SBI} we expect that if we do not discard alpha rich stars the thick disc velocities will moderate any trend in the mean azimuthal velocities with mean metallicities. In the model they balance the young disc trend at intermediate metallicities, because the downtrend for the younger low $\afe$ stars is compensated for by the increasing number of high asymmetric drift, high $\afe$ stars of the chemical thick disc towards lower metallicities. In the lower metallicity regime ($\feh < -1.0$) we even encounter a mildly positive slope in the model, because of some peculiarities in early disc formation that will be discussed in an upcoming paper. In this light it is essential to get a clean cut in the metallicity plane, removing the high $\afe$ population. The following exercise was first presented at the IAU assembly, but concerning the need for re-interpretation it seems appropriate to show it again.

\figref{fig:Borkoverview} shows the metallicity plane as defined by $\mgfe$ as alpha element against $\feh$ in the \cite{Borkova05} sample. This sample is a homogenised compilation of high-resolution spectroscopic data from several studies of local stars \citep[e.g.][]{Bensby05, Fuhrmann04, Reddy03, Reddy06}. One should keep in mind a major caveat against over-interpretation of any data we use here: Almost all local high-resolution data have strong kinematic selection biases for increasing or decreasing the number of selected thin/thick disc stars. It should also be kept in mind that because of the strong correlations between single velocity components and also between metallicities and kinematics, any such sample is uncontrollably biased both in metallicities and in velocity space. Yet, it is still interesting to look at these trends, which are confirmed on GCS data in these proceedings by L. Casagrande. As advocated by the depopulated region at intermediate $\mgfe$ we cut the sample to keep only objects with $\mgfe < 0.18$, to get a chemical thin disc selection. In light of the observational errors there might be some residual contamination, which seems to be confirmed when we measure the gradient of metallicity against angular momentum: Since the local azimuthal velocities can be translated directly into the angular momentum of those stars and let us this way estimate the stellar guiding centre radii, a local sample can be used as a good indicator of the radial abundance gradient. For our cut we estimate $d\feh/dR = -0.04 \dex/\kpc$ a value that rises up to $-0.1 \dex/ \kpc$ depending on how much of the low angular momentum regime we remove from the sample and how far down in $\mgfe$ we move the cut (the radial gradient increases mildly for stricter selections which indicates that we are tossing out some residual old stars). Interestingly the gradient for the cleaner selections is a bit higher than in the \cite{Luck11} data, but this should not be taken too seriously in the light of the diverse biases in our sample.

\begin{figure}
\begin{center}
\epsfig{file=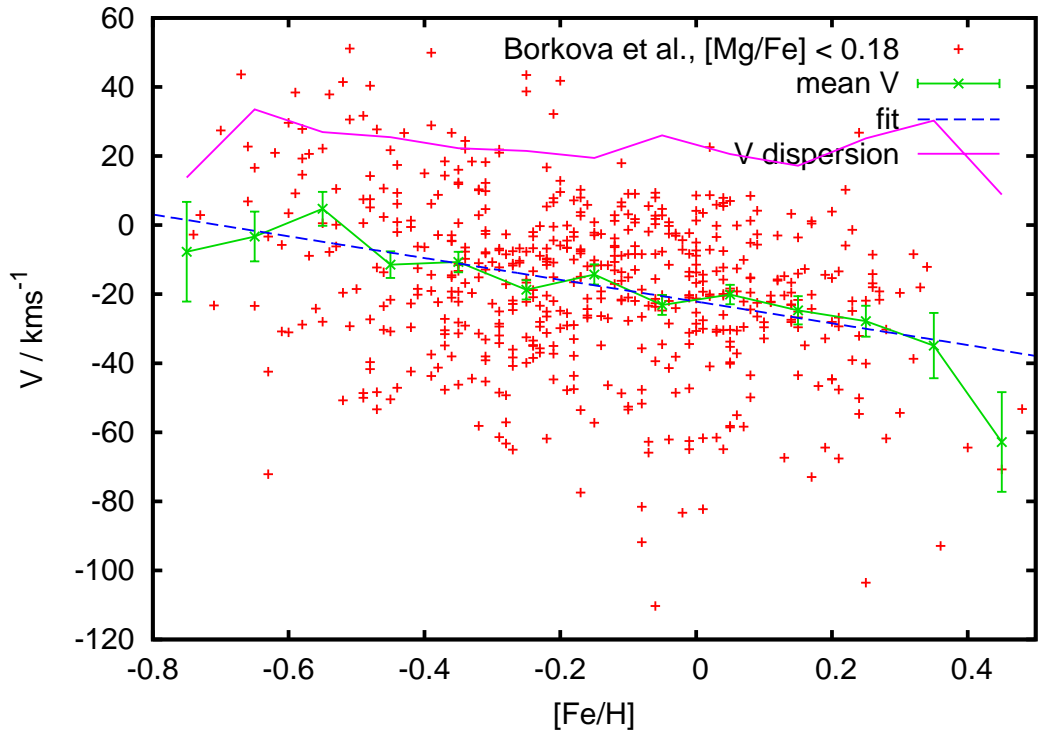, angle=-90, width=\hsize}
\caption[Kinematics versus metallicities in the Borkova et al. thin disc]{Kinematics versus metallicities in the Borkova dataset in the subsample of stars with $\mgfe < 0.18$ (red data points). The blue line gives the linear fit of the sample compared to the binned heliocentric mean velocities shown in green. The purple line shows the velocity dispersions.}\label{fig:BorkV}
\end{center}
\end{figure}

The trend of mean heliocentric $V$ velocity against metallicity in the low $\mgfe$ subsample is examined in \figref{fig:BorkV}. A linear fit (green line) yields a highly significant slope of $dV_\phi/d\feh = (-31.5 \pm 4.2) \kms / \dex$ and is plotted onto the data. The observed trend is a bit higher than what was found by \cite{Lee11} -- this can most likely be traced to the larger $\afe$ errors of the SEGUE \citep[][]{Yanny09} sample with its low resolution spectra that likely results in larger contamination of the low $\afe$ part with high $\afe$ stars. Surprisingly our finding is not consistent with the result of \cite{Navarro11} on essentially the same data. The same result as with the linear fit is seen from the binned means, where we divided the sample into $0.1 \dex$ wide subsets. At the same time we plot the dispersions for each bin (purple line). We can compare this with the qualitative predictions of the different chemical evolution models: In classical chemical evolution models \citep[e.g.][]{Chiappini01} the thin disc density ridge is created by the local population running along this ridge from low to high metallicities. This implies a significant age trend with metallicity and hence via the age-dispersion relation the most metal poor stars should show the highest velocity dispersions with a clear downtrend towards the metal rich objects. This is not observed, while the data are consistent with the behaviour in the radial migration models that favour dispersions being rather flat with metallicity at low $\afe$. Similarly the mean V velocity trend is hard to explain in the classical framework, but straight forward in the framework of radial migration models. Yet there is a little problem: The trend of V velocities with metallicity is considerably higher in the model (predicting up to $50 \kms / \dex$) than it is in the data. 

When we look at the above implications of the velocity gradient on metallicities, it becomes clear what went wrong in \cite{SBI}. When we set the radial abundance gradient in the model, we had to rely on older abundance gradient data that now seem to have indicated a too steep abundance gradient. At least it is significantly steeper than the value derived by \cite{Luck11}. The overestimated gradient places one of the main findings of the first paper on firm grounds, namely that radial migration is necessary for explaining the local metallicity distribution. Yet, it appears that with the too steep gradient we underestimated the need for stellar radial migration: The lower the gradient, the farther stars need to migrate in order to yield the same width of the local metallicity distribution. Strengthened migration should then result in a shallower slope of velocities with metallicity. As a side effect the stronger migration would result in a locally stronger and larger scale-height thick disc. The standard model from \cite{SBI} was already rather on the upper edge on how strong the thick disc may be. Again things fit together as on the decision between vertical energy and vertical action as conserved quantity \cite{SBI} decided for vertical energy. Solway et al. (in prep) and also the recent paper of \cite{Bird11} showed that vertical action is rather conserved than vertical energy. This will deliver some reduction in thick disc scale heights that can balance the stronger thick disc arising from increased radial migration. 

An impediment to modelling has so far been that despite its advantages the classic adiabatic approximation and also adiabatic modelling of stellar populations as put forward in \cite{B10} violate total energy conservation. The main effect of this short-coming is a significant underestimate of the Galactocentric radii populated by an orbit or in other words an under-prediction of inner disc stars in the intermediate and outer disc regions. Two solutions to this problem have been proposed: angular momentum correction by \cite{BM11} and the adiabatic potential of \cite{SBIII}, which directly restores total energy conservation by correcting the  horizontal potential for the energy removed from the vertical motion. This solution pushes the high vertical energy orbits back to the outer disc, compensating locally for some of the difference between isothermal approximations and the adiabatic approximation. We are currently updating the \cite{SBI} model to include the adiabatic potential. A definitive treatment of the $([Fe/H],V_\phi)$ relation must await completion of this project.

\begin{figure}
\begin{center}
\epsfig{file=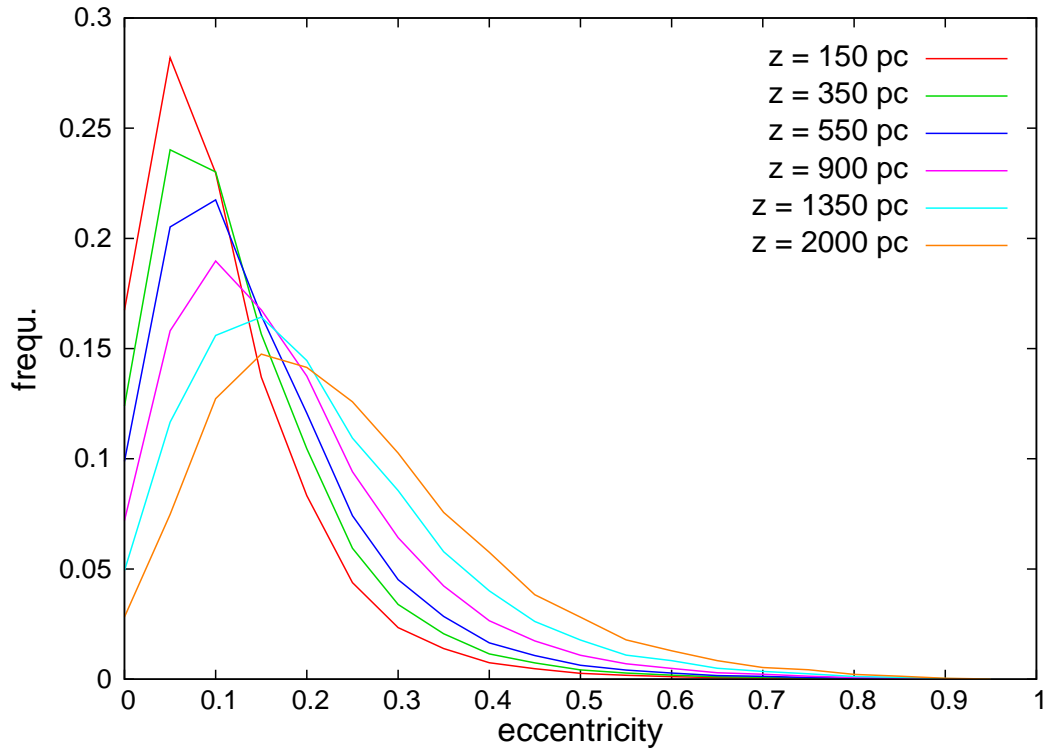, angle=-90, width=\hsize}
\caption[Eccentricity distributions from the standard model at different altitudes]{The eccentricity distribution from \cite{SBI} at different altitudes above the plane at Solar Galactocentric radius. We used simple population masses, i.e. no selection function was applied.}
\label{fig:eccdist}
\end{center}
\end{figure}  

\section{About eccentricity distributions}

In recent studies starting with \cite{Sales09} and e.g. later \cite{Lee11} the comparison of eccentricity distributions has become a frequently used tool to assess possible scenarios for the history of the Galactic disc. \figref{fig:eccdist} shows the eccentricity distribution from the \cite{SBI} model without change of any parameter and applying no selection function, i.e. using the population masses as weight. The result looks quite similar to \cite{Lee11}, yet we do not want to book this as a win, as we would like to advocate against the use of eccentricity distributions for several reasons: First it is counter-intuitive to fold the wealth of information we have into a single quantity that hides essential information like the angular momentum distribution of the disc. As another point one needs to perform a real propagation of distance, proper motion and line-of-sight velocity errors that to date does not appear to have been fully carried out. This would be computationally expensive, because in contrast to simple velocity space errors have to be calculated on the orbital model from which the eccentricities must be derived. More important the derived eccentricity values depend on the assumed potential, which governs the orbit extension of a star derived from its estimated position and velocity. Another Achilles' heel of the method -- that comparisons of eccentricities utilise a different potential for the calculation of orbits than the theoretical models they compare to -- has in most cases been neglected. One may of course argue that eccentricities express the general circularity of orbits in the different approaches. To some part this is true, but there we come to a far more serious problem: The heating in models and even principal heating mechanisms are very weakly constrained \citep[see e.g. the discussion in][]{AB09}. Locally the eccentricities are quite directly related to azimuthal velocities in that high asymmetric drifts or respectively low rotational velocities imply large eccentricities. So we see the large uncertainty in the expected eccentricities by looking at the major changes in Fig. 2 of \cite{SBIII} that are induced by moderate variation in the assumed parameters of the disc. For example heating up the inner disc of the Galaxy a bit more, which is covered either by heating from a bar or respectively simply in the uncertainties intrinsic to molecular clouds as source of random energy, the number of high eccentricity visitors from the inner disc increases significantly. This would in the one-dimensional comparison be nearly indistinguishable from high eccentricity stars contributed e.g. by a minor merger in the outer regions of the disc. Summed up there may be physical entities (e.g. domination of the Solar neighbourhood by debris from some satellite) that can be ruled out by analysis of eccentricity distributions (though it is questionable why this should not be possible on full velocity space as well). However, we see that a large range of eccentricity distributions can be explained just by uncertainty in heating and other Galactic parameters, and moreover the use of eccentricity distributions in isolation hides a lot of valuable information.

\section{Conclusions}

Analysis of local spectroscopic samples reveals for the chemically young disc a significant downtrend of mean azimuthal velocities $V_\phi$ with metallicity, while there appears to be no notable change in velocity dispersion. Those samples have to be used with caution as they carry strong kinematic biases, but the observations are confirmed by kinematically unbiased data from \cite{CSA11}. While the qualitative picture is nicely consistent with predictions of radial migration models and has so far not been explained in the context of classical approaches, we stress that in particular the azimuthal velocity trend is not a consequence of radial migration of stars by itself. On the contrary the velocity trend is a natural consequence of the radial abundance gradient in the galactic disc. Radial mixing expands the baseline in metallicity on which the trend is observable, but moderates its slope. The fact that the observed slope is a bit smaller than predictions from the \cite{SBI} model can be reasoned by an underestimate of the amount of mixing that was caused by a too steep radial abundance gradient adopted in that approach.

It is a pleasure to thank James Binney for helpful comments.


\begin{thebibliography}{}

\bibitem[Aumer \& Binney(2009)]{AB09}
Aumer M., Binney J., 2009, MNRAS, \textbf{397}, 1286

\bibitem[Bensby et al.(2005)]{Bensby05}
Bensby T., Feltzing S., Lundstr\"om I., Ilyin I., 2005, A\&A, \textbf{433}, 185

\bibitem[Binney(2010)]{B10}
Binney J., 2010, MNRAS, \textbf{401}, 2318

\bibitem[Binney \& McMillan(2011)]{BM11}
Binney J., McMillan P., 2011, MNRAS, 413, 1889 (BM11)

\bibitem[Bird et al.(2011)]{Bird11}
Bird J.C., Kazantzidis S., Weinberg D.H., 2011, MNRAS accepted, arXiv:1104.0933

\bibitem[Borkova \& Marsakov(2005)]{Borkova05}
Borkova T.V., Marsakov V.A., 2005, ARep, \textbf{49}, 405

\bibitem[Casagrande et al.(2011)]{CSA11}
Casagrande L., Sch\"onrich R., Asplund M., Cassisi S., Ram\'irez I.,  Mel\'endez J., Bensby T., Feltzing S., 2011, A\&A, \textbf{530}, 138

\bibitem[Chiappini et al.(2001)]{Chiappini01}
Chiappini C., Matteucci F., Romano D., 2001, ApJ, \textbf{554}, 1044

\bibitem[Fuhrmann(2004)]{Fuhrmann04}
Fuhrmann K., 2004, AN, \textbf{325}, 3 

\bibitem[Haywood(2008)] {Haywood08}
Haywood M., 2008, MNRAS, 388, 1175

\bibitem[Lee et al.(2011)]{Lee11}
Lee Y.S. et al., 2011, ApJ, \textbf{738}, 187

\bibitem[Loebman et al.(2011)]{Loebman11}
Loebman S.R., Ro{\v s}kar R., Debattista V.P., Ivezi{\'c} {\v Z}., Quinn T.R., Wadsley J., 2011, ApJ, 737, 8

\bibitem[Luck \& Lambert(2011)]{Luck11}
Luck R.E., Lambert D.L., 2011, AJ, \textbf{142}, 136

\bibitem[Matteucci \& Brocato(1990)]{Matteucci90}
Matteucci F., Brocato E., 1990, ApJ, 365, 539

\bibitem[Navarro et al.(2011)]{Navarro11}
Navarro J.F., Abadi M.G., Venn K.A., Freeman K.C., Anguiano B., 2011, MNRAS, \textbf{412}, 1203

\bibitem[Reddy et al.(2003)]{Reddy03}
Reddy B.E., Tomkin J., Lambert D.L., Allende Prieto C., 2003, MNRAS, \textbf{340}, 304

\bibitem[Reddy et al.(2006)]{Reddy06}
Reddy B.E., Lambert D.L., Allende Prieto C., 2006, MNRAS, \textbf{367}, 1329

\bibitem[Sales et al.(2009)]{Sales09}
Sales L.V., 2009, MNRAS, \textbf{400}, 61

\bibitem[Sch\"onrich \& Binney(2009a)]{SBI}
Sch\"onrich R., Binney J., 2009a, MNRAS, \textbf{396}, 203

\bibitem[Sch\"onrich \& Binney(2009b)]{SBII}
Sch\"onrich R., Binney J., 2009b, MNRAS, \textbf{399}, 1145

\bibitem[Sch\"onrich \& Binney(2011)]{SBIII}
Sch\"onrich R., Binney J., 2011, MNRAS, accepted, arXiv: 1109.4417

\bibitem[Yanny et al.(2009)]{Yanny09}
Yanny B. et al., 2009, ApJ, 137, 4377

\end{thebibliography}
\end{document}